%
% the following is to use blackboard bold fonts --
\let\useblackboard=\iftrue
%
% activate this if you don't have them.
%\let\useblackboard=\iffalse
%
% You might also need to remove this line.
\newfam\black
\input harvmac.tex
%\input tables.tex
%\input labeldefs.tmp
%labeldefs.tmp
\def\dhn{[1]}
\def\bfss{[2]}
\def\seiberg{[3]}
\def\bilal{[4]}
\def\wati{[5]}
\def\yoneya{[6]}
\def\dewit{[7]}
\def\ezawa{[8]}
\def\memlag{(2.1)}
\def\apddef{(2.2)}
\def\memham{(2.3)}
\def\membra{(2.4)}
\def\memglaw{(2.5)}
\def\memxmin{(2.6)}
\def\memxm{(2.7)}
\def\memsusy{(2.8)}
\def\memlor{(2.9)}
\def\sphhar{(2.10)}
\def\sphorth{(2.11)}
\def\sphcom{(2.12)}
\def\sphcomt{(2.13)}
\def\sphgr{(2.14)}
\def\sphten{(2.15)}
\def\sphgaus{(2.16)}
\def\comrel{(2.17)}
\def\sphham{(2.18)}
\def\xiym{(2.19)}
\def\lorgen{(2.20)}
\def\wedefi{(2.21)}
\def\marquard{[9]}
\def\sethi{[10]}
\def\mambor{(3.1)}
\def\orterms{(3.2)}
\def\toneo{(3.3)}
\def\eigtwo{(3.4)}
\def\secter{(3.5)}
\def\seclef{(3.6)}
\def\eqeig{(3.7)}
\def\ftheta{(3.8)}
\def\fftheta{(3.9)}
\def\fthen{(3.10)}
\def\eigsix{(3.11)}
\def\ninfou{(3.12)}
\def\nineig{(3.13)}
\def\onehun{(3.14)}
\def\mahone{(3.15)}
\def\mahfer{(3.16)}
\def\tqter{(3.17)}
\def\susskind{[11]}
\def\brezin{[12]}
\def\sakai{[13]}
\def\douglas{[14]}
\def\renorm{(4.1)}
\def\simren{(4.2)}
\def\simssren{(4.3)}
\def\bfssqm{(4.4)}
\def\jevicki{[15]}
\def\dmnid#1{\hbox {$(\hbox {A.}1#1)$}}
\def\idstuff{(\hbox {A.}2)}
\def\ezaid#1{\hbox {$(\hbox {A.}3#1)$}}
\def\gambas{(\hbox {B.}1)}
\def\stiff{(\hbox {B.}2)}
\def\gammid{(\hbox {B.}3)}
\def\gammidd{(\hbox {B.}4)}
\def\ffermid{(\hbox {B.}5)}
%
%%%%%%%%%%%%%%%%%%%%%%%%%%%%%%%%%%%%%%%%%%%%%%%%%%%%%%%%%%%%%%%
%The following lines are needed to insert the accompanying figures in
%the paper. If you do not have epsf, then comment out the line
% ``\input epsf'', and print the figures separately. The figures are
%at
%the end of the tex file, with instructions for their extraction.
%\input epsf.tex
\ifx\epsfbox\UnDeFiNeD\message{(NO epsf.tex, FIGURES WILL BE
IGNORED)}
\def\figin#1{\vskip2in}% blank space instead
\else\message{(FIGURES WILL BE INCLUDED)}\def\figin#1{#1}\fi
\def\ifig#1#2#3{\xdef#1{fig.~\the\figno}
\midinsert{\centerline{\figin{#3}}%
\smallskip\centerline{\vbox{\baselineskip12pt
\advance\hsize by -1truein\noindent{\bf Fig.~\the\figno:} #2}}
\bigskip}\endinsert\global\advance\figno by1}
%%%%%%%%%%%%%%%%%%%%%%%%%%%%%%%%%%%%%%%%%%%%%%%%%%%%%%%%%%%%%%%%
\noblackbox
\def\Title#1#2{\rightline{#1}
\ifx\answ\bigans\nopagenumbers\pageno0\vskip1in%
\baselineskip 15pt plus 1pt minus 1pt
\else%\special{papersize=11in,8.5in}%
\def\listrefs{\footatend\vskip
1in\immediate\closeout\rfile\writestoppt
\baselineskip=14pt\centerline{{\bf
References}}\bigskip{\frenchspacing%
\parindent=20pt\escapechar=` \input
refs.tmp\vfill\eject}\nonfrenchspacing}
\pageno1\vskip.8in\fi \centerline{\titlefont #2}\vskip .5in}
 
scaled\magstep3
 
scaled\magstep3
 
scaled\magstep3
 
scaled\magstep3
 
scaled\magstep3
\ifx\answ\bigans\def\tcbreak#1{}\else\def\tcbreak#1{\cr&{#1}}\fi

\useblackboard
\message{If you do not have msbm (blackboard bold) fonts,}
\message{change the option at the top of the tex file.}

\font\blackboard=msbm10 scaled \magstep1
\font\blackboards=msbm7
\font\blackboardss=msbm5
%\newfam\black
\textfont\black=\blackboard
\scriptfont\black=\blackboards
\scriptscriptfont\black=\blackboardss

\else

\fi
% *************************************
%\draftmode
%

%
\def\yboxit#1#2{\vbox{\hrule height #1 \hbox{\vrule width #1
\vbox{#2}\vrule width #1 }\hrule height #1 }}
\def\fillbox#1{\hbox to #1{\vbox to #1{\vfil}\hfil}}
\def\ybox{{\lower 1.3pt \yboxit{0.4pt}{\fillbox{8pt}}\hskip-0.2pt}}

\def\comments#1{}

\def\half{{1\over 2}}
\def\Tr{{{\rm Tr\  }}}

\def\CF{{\cal F}}

\def\CM{{\cal M}}

\def\CL{{\cal L}}

\def\II{\relax{I\kern-.07em I}}
\def\IIA{{\II}A}

\def\inbar{\,\vrule height1.5ex width.4pt depth0pt}
\def\IZ{\relax\ifmmode\mathchoice
{\hbox{\cmss Z\kern-.4em Z}}{\hbox{\cmss Z\kern-.4em Z}}
{\lower.9pt\hbox{\cmsss Z\kern-.4em Z}}
{\lower1.2pt\hbox{\cmsss Z\kern-.4em Z}}\else{\cmss Z\kern-.4em
Z}\fi}
\def\IB{\relax{\rm I\kern-.18em B}}
\def\IC{{\relax\hbox{$\inbar\kern-.3em{\rm C}$}}}
\def\ID{\relax{\rm I\kern-.18em D}}
\def\IE{\relax{\rm I\kern-.18em E}}
\def\IF{\relax{\rm I\kern-.18em F}}
\def\IG{\relax\hbox{$\inbar\kern-.3em{\rm G}$}}
\def\IGa{\relax\hbox{${\rm I}\kern-.18em\Gamma$}}
\def\IH{\relax{\rm I\kern-.18em H}}
\def\IK{\relax{\rm I\kern-.18em K}}
\def\IP{\relax{\rm I\kern-.18em P}}
\def\pp{{\relax{=\kern-.42em |\kern+.2em}}}
%\def\IX{\relax{\rm X\kern-.01em X}}
%this doesn't work

\font\cmss=cmss10 \font\cmsss=cmss10 at 7pt
\def\IR{\relax{\rm I\kern-.18em R}}

\def\Tr{{\rm Tr\ }}

\def\frac#1#2{{{#1} \over {#2}}}

%
%Journal macros
%

\def\NP{{\it Nucl. Phys.\ }}

\def\PL{{\it Phys. Lett.\ }}
\def\PR{{\it Phys. Rev.\ }}
\def\PRL{{\it Phys. Rev. Lett.\ }}
\def\CMP{{\it Comm. Math. Phys.\ }}

\def\ATMP{{\it ATMP\ }}
\writedefs

\Title{ \vbox{\baselineskip12pt\hbox{hep-th/9807229}
\hbox{BROWN-HET-1133}
}}
{\vbox{
\centerline{Eleven-Dimensional Lorentz Symmetry}
\centerline{ from SUSY Quantum Mechanics}}}

\centerline{ David A. Lowe}
\medskip

\centerline{Department of Physics}
\centerline{Brown University}
\centerline{Providence, RI 02912, USA}
\centerline{\tt lowe@het.brown.edu}
\bigskip

\centerline{\bf{Abstract}}

The supermembrane in light-cone gauge gives rise to a supersymmetric
quantum mechanics system with $SU(N)$ gauge symmetry when the group of
area preserving diffeomorphisms is suitably regulated. de Wit,
Marquard and Nicolai showed how eleven-dimensional Lorentz generators
can be constructed from these degrees of freedom at the classical
level. In this paper, these considerations are extended to the quantum
level and it is shown the algebra closes to leading nontrivial
order at large $N$. A proposal is made for extending these
results to Matrix theory by realizing longitudinal boosts 
as large $N$ renormalization group transformations.

\vfill
\Date{\vbox{\hbox{\sl July, 1998}}}
%References
\lref\susskind{L. Susskind, ``Another Conjecture About M(atrix)
Theory,'' hep-th/9704080.}
\lref\jevicki{A. Jevicki, ``Light-Front Partons and Dimensional
Reduction in Relativistic Field Theory,'' \PR {\bf D57} (1998) 5955,
hep-th/9712088.} 
\lref\wati{W. Taylor, ``Lectures on D-branes, Gauge Theory and
M(atrices),'' hep-th/9801182.}
\lref\bfss{T. Banks, W. Fischler, S.H. Shenker and L. Susskind, ``M
Theory as a Matrix Model: A Conjecture,'' hep-th/9610043.}
\lref\dewit{B. de Wit, U. Marquard and H. Nicolai, ``Area Preserving
Diffeomorphisms and Supermembrane Lorentz Invariance,'' \CMP {\bf 128}
(1990) 39.}
\lref\marquard{U. Marquard and M. Scholl, ``Lorentz Algebra And
Critical Dimension For The Bosonic Membrane,'' \PL {\bf B227} (1989)
227;
U. Marquard, R. Kaiser and M. Scholl, ``Lorentz Algebra And Critical 
Dimension For The Supermembrane,'' \PL {\bf B227} (1989) 234.}
\lref\dhn{B. de Wit, J. Hoppe and H. Nicolai, ``On the quantum
mechanics of
supermembranes,'' \NP {\bf B305} (1988) 545.}
\lref\ezawa{K. Ezawa, Y. Matsuo and K. Murakami, ``Lorentz Symmetry of
Supermembrane in Light Cone Gauge Formulation,'' hep-th/9705005.}
\lref\brezin{E. Brezin and J. Zinn-Justin, ``Renormalization Group
Approach to Matrix Models,'' hep-th/9206035.}
\lref\sakai{ S. Higuchi, C. Itoi, S. Nishigaki and N. Sakai,
``Renormalization Group Flow In One And Two Matrix Models,'' \NP {\bf
B434} (1995) 283, hep-th/9409009.}
\lref\douglas{M.R. Douglas, ``D-branes and Matrix Theory in Curved
Space,'' hep-th/9707228.}
\lref\seiberg{A. Sen, ``D0-Branes On T**N And Matrix Theory,''
\ATMP {\bf 2} (1998) 51, hep-th/9709220;
N. Seiberg, ``Why Is The Matrix Model Correct?''
\PRL {\bf 79} (1997) 3577, hep-th/9710009.}
\lref\yoneya{Y. Okawa, T. Yoneya, 
``Multibody Interactions Of D Particles In Supergravity And
Matrix Theory,'' hep-th/9806108.}
\lref\pouliot{J. Polchinski and P. Pouliot, ``Membrane Scattering With
M Momentum Transfer,'' \PR {\bf D56} (1997) 6601, hep-th/9704029.}
\lref\sethi{S. Sethi and Mark Stern, ``D-Brane Bound States Redux,''
\CMP {\bf 194} (1998) 675, hep-th/9705046.}
\lref\bilal{V. Balasubramanian, R. Gopakumar and F. Larsen, ``Gauge
Theory, Geometry and the Large N Limit,'' hep-th/9712077;
A. Bilal, ``A Comment on Compactification of M-Theory on
an (Almost) Light-Like Circle,'' \NP {\bf B521} (1998) 202, hep-th/9801047;
``DLCQ of M-theory as the light-like limit,''
hep-th/9805070; A. Guijosa, ``Is Physics in the Infinite Momentum
Frame Independent of the Compactification Radius?'' hep-th/9804034.} 

\newsec{Introduction}

Some time ago, de Wit, Hoppe and Nicolai \dhn\ showed that the
supermembrane in light-cone gauge reduces to maximally supersymmetric quantum
mechanics with $SU(N)$ gauge symmetry in the large $N$ limit. 
The $SU(N)$ gauge
symmetry arises as a regulated version of the area preserving
diffeomorphism symmetry of the supermembrane.
More recently \bfss\
Banks, Fischler, Shenker and Susskind 
proposed to use this supersymmetric quantum mechanics as a 
complete description of eleven-dimensional M-theory. This description
has come to be known as Matrix theory.
The key
distinction between this new point of view and the previous work is
the identification of $N$ with the number of quanta of 
momentum along a compact direction. The relationship between
supersymmetric quantum mechanics and M-theory was clarified in
\seiberg. There it was argued that if one assumes eleven-dimensional
Lorentz invariance, and the duality between Type \IIA\ string theory
and M-theory on a spacelike circle, 
M-theory on a light-like circle reduces to the
supersymmetric quantum mechanics system. The limiting procedure
needed for this argument to work has been further studied in \bilal.

In this paper we will take a different point of view and study how
eleven-dimensional Lorentz invariance can be recovered, starting with
the supersymmetric quantum mechanics.
%In this paper we will study the construction of the
%eleven-dimensional Lorentz algebra in terms of the large
%$N$ limit of the supersymmetric quantum mechanics. 
A number of 
scattering amplitude calculations already provide evidence for the
Lorentz invariance of Matrix theory at large $N$.
See the review \wati\ for a discussion of some of these results
and further references. A notable recent example was the computation
using Matrix theory of 
the three graviton scattering amplitude, in
agreement with the prediction of supergravity \yoneya.

Ideally one would like
a direct construction of the Lorentz generators 
using the quantum mechanics degrees of freedom.  
In the original dWHN \dhn\ 
interpretation of the supersymmetric quantum mechanics, these Lorentz
generators where constructed in \dewit\ and shown to close at the
classical level, up to $1/N^2$ corrections \refs{\dewit,\ezawa}. On
the other hand,
for finite $N$ Matrix theory the construction of such generators
seems difficult since longitudinal
boosts change the value of $N$. Suitable 
definitions should exist for large $N$, as will be discussed 
in the following.

In this paper we extend the analysis of the dWHN model 
to the quantum theory by
properly dealing with ordering ambiguities, and find the Lorentz
algebra closes to leading nontrivial order at large $N$.
We propose a way to extend these results to Matrix theory, by 
realizing longitudinal boosts as large $N$ renormalization 
group transformations.

\newsec{Supermembranes and SUSY Quantum Mechanics}

In this section we review the relationship between the light-cone
formulation of the supermembrane and supersymmetric quantum mechanics
\dhn. For the most part we will follow the notation of \dewit.
After gauge fixing, the Lagrangian takes the form
\eqn\memlag{
{1\over \sqrt{w}}\CL = \frac{1}{2}(D_0 {\vec X})^2
+\frac{i}{2}\theta D_0 \theta -\frac{1}{4}
\left(\left\{X^a, X^b\right\}\right)^2
+\frac{i}{2}\theta\gamma_a\left\{X^a,\theta\right\}~,
}
where
$X^a(t,\sigma^r)$, $\theta_\alpha(t,\sigma^r)$
($a=1,\ldots,9$, $\alpha=1,\ldots,16$, $r=1,2$)
are the transverse worldvolume
degrees of freedom dependent on the worldvolume coordinates $t$ and
$\sigma^r$.
The indices $a$ and $\alpha$ are respectively
the vector and the spinor degrees of freedom of $SO(9)$. The
conventions for gamma matrices are described in Appendix B.
$w_{ij}$ is the $2\times 2$
spatial metric tensor on the worldvolume
and $w$ is its determinant.
The bracket is defined as,
$
\left\{A,B\right\}
\equiv \frac{\epsilon^{rs}}{\sqrt{w(\sigma)}}
\partial_r A(\sigma) \partial_s B(\sigma)~.
$
The covariant derivative,
$
D_0 X^a=\partial_0 X^a-\left\{\omega,X^a\right\}
$,
$
D_0 \theta=\partial_0 \theta-\left\{\omega,\theta\right\},
$
defines the gauge transformation corresponding to an area
preserving diffeomorphism (APD),
\eqn\apddef{
\delta X^a=\left\{\xi, X^a\right\},\quad
\delta \theta=\left\{\xi, \theta\right\},\quad
\delta\omega=\partial_0\xi+\left\{\xi, \omega\right\}~.
}

The canonical Hamiltonian  is
\eqn\memham{
\eqalign{
H &= -\int d^2 \sigma P^{-}(\sigma)\cr
&= \frac{1}{P^+_0}
\int d^2 \sigma
\sqrt{w(\sigma)}
\left(
\frac{1}{2}w^{-1}\vec{P}^2+\frac{1}{4}
\left(\left\{X^a,X^b\right\}\right)^2
{} -\frac{i}{2}\theta\gamma_a\left\{X^a,\theta\right\}
\right)~.\cr}
}
$\vec{P}$ denotes the canonical momentum conjugate to
$\vec{X}$.
The non-vanishing Dirac brackets are
\eqn\membra{
\eqalign{
\left(X^a(\sigma),P^b(\rho)\right)_{DB}
& =  \delta^{ab}\delta^{(2)}(\sigma,\rho),\cr
\left(\theta_\alpha(\sigma),\theta_\beta(\rho)\right)_{DB}
&= -\frac{i}{\sqrt{w(\sigma)}}\delta_{\alpha\beta}
\delta^{(2)}(\sigma,\rho)~.\cr}
}

The Gauss law constraints associated with the APD
are
\eqn\memglaw{
\varphi(\sigma) \equiv
{} - \left\{ \frac{P^a(\sigma)}{w(\sigma)},
{X^a}(\sigma)\right\}
{} - \frac{i}{2}\left\{
\theta(\sigma),\theta(\sigma)\right\}\approx 0~.
}
Here we take the membrane to have spherical topology. For higher
topology there are additional APD's generated by the harmonic vectors,
which give rise to additional Gauss law constraints. 

The light cone directions are defined as
$X^\pm=\frac{1}{\sqrt{2}}(X^{10}\pm X^0)$ where
$X^+$ is related to the worldvolume time $t$ by
$X^+(\tau)= X^+(0)+t$, and
\eqn\memxmin{
\partial_r X^-(\sigma)
=
{} -\frac{1}{P^+_0}\left(
\frac{1}{\sqrt{w(\sigma)}}
\vec{P}(\sigma)\cdot\partial_r\vec{X}(\sigma)
+\frac{i}{2}\theta(\sigma)\partial_r\theta(\sigma)\right)~.
}
The integrability conditions of this differential equation
coincide with the Gauss law constraints. When integrated, it gives
\eqn\memxm{
X^{-}(\sigma) =
q^- - \frac{1}{P^+_0}\int d^2 \rho G^{r}(\sigma, \rho)
\left( \vec{P}(\rho)\cdot \partial_r\vec{X}(\rho) +
\frac{i}{2}\sqrt{w(\rho)} \theta(\rho)\partial_r \theta(\rho)
\right)~,
}
where the integration constant
satisfies $(q^-,P^+_0)_{DB}=1$ and
$G^r(\sigma, \rho)$ is the Green function
defined by
$D^{\rho}_rG^r(\sigma,\rho) = -(w(\sigma))^{-1/2}
\delta^{(2)}(\sigma, \rho) +1
$. 

This system has supersymmetry generated by
\eqn\memsusy{
\eqalign{
Q^+ & =  \frac{1}{\sqrt{P^+_0}}\int d^2\sigma\left(
P^a\gamma_a+ \frac{\sqrt{w}}{2} \left\{
X^a, X^b\right\} \gamma_{ab} \right)
\theta,\cr
Q^- & =  \sqrt{P_0^+}\int d^2\sigma \sqrt{w(\sigma)}
\theta~. \cr}
}

The Lorentz generators are defined by
\eqn\memlor{
\eqalign{
M^{ab} & =  \int d^2\sigma
\left( -P^aX^b + P^b X^a - \frac{i}{4} \theta \gamma^{ab}
\theta\right),\cr
M^{+-} & =  \int d^2 \sigma \left(
{} -P^+X^- + P^- X^+\right),\cr
M^{+a} & =  \int d^2 \sigma (-P^+ X^a + P^a X^+),
\cr
M^{-a} & =  \int d^2\sigma \left(
P^a X^- - P^-X^a - \frac{i}{4P^+_0}
\theta\gamma^{ab}\theta P_b
{} - \frac{i\sqrt{w}}{8P^+_0}
\left\{ X_b, X_c\right\} \theta \gamma^{abc}
\theta \right)~.\cr}
}

To regulate this theory we follow \dhn, and expand all the worldvolume
fields in terms of
spherical harmonics, with a mode cutoff dependent on $N$. The
eigenfunctions satisfy
\eqn\sphhar{
\Delta Y_0 = 0, \quad
\Delta Y_A = -\omega_A Y_A~,
}
with $\omega_A >0$. Indices $A,B,C$ are positive integers, $I,J,K$
denote non-negative integers.
The $Y_I$ are orthonormal
\eqn\sphorth{
\int d^2 \sigma \sqrt{w(\sigma)}
Y^I(\sigma) Y_J(\sigma) = {\delta^I}_{J}
\qquad
Y^I\equiv Y_I^* = \eta^{IJ}Y_J~.
}
The completeness relations take the form
\eqn\sphcom{
\sum_{A} Y^A(\sigma) Y_A(\rho) =
\frac{1}{\sqrt{w(\sigma)}}\delta^{(2)}(\sigma ,\rho)
{} -1~,}
and
\eqn\sphcomt{
\eqalign{
& \sum_{A}\frac{1}{\omega_{A}}
     [ D^{r}Y_{A}(\sigma)D^{s}Y^{A}(\rho)
      + \frac{\epsilon^{rt}}{\sqrt{w(\sigma)}}\partial_{t}Y_{A}(\sigma)
           \frac{\epsilon^{su}}{\sqrt{w(\rho)}}\partial_{u}Y^{A}(\rho)
    ] \cr
    & \qquad \qquad
=\frac{w^{rs}(\sigma)}{\sqrt{w(\sigma)}}\delta^{(2)}(\sigma,
    \rho)~.}
}
The Green function is rewritten,
\eqn\sphgr{
G^r(\sigma, \rho) = \sum_A
\frac{1}{\omega_A}Y^A(\sigma) \partial^r Y_A(\rho)~.
}
Three-index tensors that will be used in the following are:
\eqn\sphten{
\eqalign{
f_{ABC} &= \int d^2\sigma
\sqrt{w(\sigma)} Y_A(\sigma) \left\{
Y_B(\sigma),Y_A(\sigma) \right\}
\cr
d_{ABC} & = \int d^2\sigma
\sqrt{w(\sigma)} Y_A(\sigma)Y_B(\sigma)Y_C(\sigma)\cr
c_{ABC} & =
{} -2\int d^2\sigma
\sqrt{w(\sigma)} \frac{w^{rs}}{\omega_A}
\partial_rY_A Y_B \partial_s Y_C~.\cr}
}
These tensors satisfy a number of nontrivial identities which are
described in Appendix A.

The approximation of the group of area preserving diffeomorphisms on
$S^2$ by the group $SU(N)$ is considered in detail in appendix B of
\dewit (see also references therein). Here we will only need to know
that $A=1,\cdots,N^2-1$ and that as $N\to \infty$ $f_{ABC}$ is well
approximated by the structure constants of $SU(N)$, $d_{ABC}$ is
proportional to the three index symmetric tensor, and $c_{ABC}$ is a
linear combination of the invariant tensors $f_{ABC}$ and $d_{ABC}$.
It should be pointed out that $c_{ABC}$ is not invariant under APD's
\dewit, since it explicitly depends on the worldvolume metric. 
However once the group of APD's is regulated via $SU(N)$ it is 
possible to modify the definition of $c_{ABC}$ so that it is invariant 
under $SU(N)$.
For concreteness we follow the definitions of \dewit, appendix B. With
this approximation 
the identities of
Appendix A are satisfied up to $1/N^2$
corrections.

Now the membrane formulae above may be rewritten in terms of $SU(N)$
variables and the zero modes.
The Gauss law constraints are re-expressed as
\eqn\sphgaus{
\varphi_A  =  f_{ABC}\left(
\vec{X}^B\cdot\vec{P}^C -\frac{i}{2}\theta^B\theta^C\right)~.
}
Expressing the Dirac brackets as commutators we have:
\eqn\comrel{
\matrix{
\hfill [X_a^A, P_{bB} ] &= & i \delta_{ab} \delta_B^A &\qquad&
\{ \theta_{\alpha }^A , \theta_{\beta B} \}  &=& \delta_{\alpha \beta}
\delta_B^A \cr
[ q^-, P_0^+ ] &=& i &\qquad&
[X_{a0}, P_{b0} ] &=& i \delta_{ab} \cr
\{ \theta_{\alpha 0} , \theta_{\beta 0} \}  &=& \delta_{\alpha \beta}
~.\cr }
}
The Hamiltonian takes the form
\eqn\sphham{
\eqalign{
H & =  \frac{\vec{P}_0^2}{2P^+_0}+\frac{
\CM^2}{2P^+_0} \cr
\CM^2 & =  \vec{P}_A^2+\frac{1}{2} (
f_{ABC}X^B_aX^C_b)^2-
if_{ABC} \theta^A \gamma^aX^B_a\theta^C ~.\cr
}}
The $X^-$ coordinate is rewritten
\eqn\xiym{
X_A^- = - {1\over P_0^+} \bigl( \vec{P}_0 \cdot \vec{X}_A +
\frac{i}{2} \theta_0 \theta_A \bigr) + \frac{1}{2 P_0^+} c_{ABC}
\bigl( \vec{P}^B \cdot \vec{X}^C + \frac{i}{2} \theta^B \theta^C
\bigr)~.
}
The Lorentz generators at $t=0$ are \dewit:
\eqn\lorgen{
\eqalign{
M^{ab} & = -P_0^a X_0^b + P_0^b X_0^a - {i\over 4} \theta_0
\gamma^{ab} \theta_0
-P^a_A X^{bA} + P^b_A X^{aA} - {i \over 4} \theta_A \gamma^{ab}
\theta^A \cr
M^{+a} &= - P_0^+ X_0^a \cr
M^{+-} &= - P_0^+ q^- + A \cr
M^{-a } &= (M^{-a})^{(0)} + {1\over P_0^+} \bigl( P_{0b} \tilde M^{ab}
-
{i \over 2} \theta_0 \gamma^a \tilde Q^+ \bigr) +
{1\over P_0^+} \tilde M^{-a} ~,\cr
}
}
where we have defined
\eqn\wedefi{
\eqalign{
(M^{-a})^{(0)} &= q^- P_0^a + X_0^a H -
{i \over 4 P_0^+} \theta_0 \gamma^{ab} \theta_0 P_{0b} + B { P_0^a
\over P_0^+}\cr
\tilde Q^+ &= ( P^a_A \gamma_a +
\half f_{ABC} X_a^B X_b^C  \gamma^{ab} )\theta^A \cr
\tilde M^{ab} &= -P_A^a X^{bA} + P^b_A X^{aA} -
{i \over 4} \theta_A \gamma^{ab} \theta^A \cr
\tilde M^{-a} &= \half d^{ABC} X_A^a ( P_B \cdot P_C + \half
(f_B^{~DE} X_D^b X_E^c)(f_C^{~FG} X_F^b X_G^c) - i f_C^{~DE} X_D^b
\theta_B \gamma_b \theta_E ) \cr
&- {i\over 4} d^{ABC} P_{Ab} \theta_B \gamma^{ab} \theta_C
+ \half c^{ABC} P_A^a ( P_B \cdot X_C + \half i \theta_B \theta_C) \cr
&- {i\over 8} f^{ABC} d_A^{~DE} X_{Bb} X_{Cc} \theta_D \gamma^{abc}
\theta_E ~.\cr}
}

We adopt an ordering prescription that preserves the $SO(9)$
rotational symmetry and take
all products of (fermionic) bosonic operators to be (a)symmetrically
ordered. In \sphham, \lorgen, \wedefi\ the possible ordering terms
have been included. The terms with coefficients $A$ and $B$ are the
only terms allowed by the symmetries.

$A$ is fixed by demanding that $[M^{+a}, M^{-a} ]= i \eta^{aa}
M^{+-}$ is satisfied. This fixes $A=0$. To fix $B$ we consider
$[M^{-a}, X_0^a] \sim q^-+ P_0^a /P_+$. To remove the
inhomogeneous term we need to set $B=i$. This follows from the
fact that $X_0^a H+ i P_0^a/P_+$ with symmetric ordering equals
$X_0^a H$ without symmetric ordering.

\newsec{Lorentz Algebra in Quantum Theory}

The Lorentz algebra has been considered at the classical level in
\refs{\dewit, \ezawa}, where it was found the algebra closes up to $1/N^2$
corrections. Typically, ordering terms appear with
extra factors of $N^2$ arising from the trace over group
generators. We will show the Lorentz algebra closes at the quantum
level up to order 1 ordering terms. To go beyond this calculation
would require a computation at the classical level to the next
nontrivial order. It is straightforward to see that the symmetries
allow nontrivial ordering terms only in the commutators $[M^{-a},
M^{-b}]$ and  $[M^{-a}, H]$. In the following we compute the ordering
terms for these commutators. The terms that have already been shown to
vanish at the classical level \refs{\dewit,\ezawa} will not be discussed
further.

One might expect that these ordering terms will only vanish in the critical
dimension of the supermembrane \marquard, as in the analogous
light-cone string theory calculation. In that case, the vanishing of
the normal ordered
commutator $[M^{-a},M^{-b}]$ led to a critical dimension of ten. Here
we will find the gauge symmetry is much more restrictive than in the
string case, and the ordering terms will vanish identically at leading
nontrivial order at large $N$ for any dimension in which the classical 
algebra closes (although our explicit calculations are for the case of 
eleven dimensions only). Certain spinor identities (see appendix 
B) are
needed to prove the closure of the algebra at the classical
level. These restrict the possible allowed dimensions to $4,5,7$ or
$11$. Of course the BFSS interpretation of the supersymmetric quantum
mechanics can only hold in eleven dimensions because only in that case 
do we get a normalizable ground state for the $N=2$ system
corresponding to a graviton with two units of longitudinal momentum \sethi.

\subsec{ $[M^{-a}, M^{-b}]$ }

The nontrivial ordering contributions come from the following terms,
which are written without (a)symmetric ordering
\eqn\mambor{
[M^{-a}, M^{-b}] = {1\over (P_0^+)^2 } (A_1+A_2+A_3+A_4)~,
}
with
\eqn\orterms{
\eqalign{
A_1 &= -{i \over 2}( \CM^2 \tilde M^{ab} +\tilde M^{ab} \CM^2) \cr
A_2 &= {1\over 4} \tilde Q^+ \gamma^{ab} \tilde Q^+ \cr
A_3 &= [ \tilde M^{-a}, \tilde M^{-b}] \cr
A_4 &= \half \theta_0 \gamma^a [ \tilde Q^+, \tilde M^{-b}] -
(a\leftrightarrow b)~. \cr}
}

It is easy to see ordering terms of order $P^2$ vanish by  
symmetry considerations alone.
Let us
ignore for the moment
the fermionic terms.
The other bosonic terms arise from ordering expressions of order $P
X^5$. We choose to order these terms so the $P$ factors are on the left.
When we move the $P$ factor to the left in \orterms,
$X^4$ terms are generated. The reordered $P X^5$ terms
combine according to the classical analysis of \ezawa. 
We must show
the extra terms generated by this reordering cancel.

The terms appearing in $A_1$ are proportional to
\eqn\toneo{
f_{ABC} f^A_{~DE} X^{Bb} X^{C}_d X^{Da} X^{Ed} - (a \leftrightarrow b)~,
}
which vanishes since the first term is symmetric under interchange of
$a$ and $b$.

To extract the ordering terms that arise in $A_3$ it is helpful to use
the analysis of \ezawa\ as much as possible. Following through the
steps in that calculation, the computation of the ordering term
amounts to ordering eqn. (82) of \ezawa\
\eqn\eigtwo{
\eqalign{
&\frac{1}{2}(f_{ABC}X_{d}^{B}X_{e}^{C})^{2}
(X^{a}_{D}P^{bD}-X^{b}_{D}P^{aD})
\cr
&+\frac{1}{2}d^{AIC}f_{I}^{\;DE}(X^{a}_{A}X^{b}_{E}-X^{b}_{A}X^{a}_{E})
\vec{X}_{D}\cdot\vec{X}_{F}c_{HC}^{\quad F}f^{HBG}\vec{X}_{G}\cdot\vec{P}_{B}
\cr
& \quad +f^{DAE}(X^{a}_{A}X^{b}_{E}-X^{b}_{A}X^{a}_{E})
\vec{X}_{D}\cdot\vec{X}_{F}f^{FBG}\vec{X}_{G}\cdot\vec{P}_{B}~.}
}
Ordering the first term gives no extra terms by the same calculation
as for $A_1$. Likewise the same tensor structures appear when ordering
the third term, so that also vanishes. The second term gives rise to the
ordering terms
\eqn\secter{
\eqalign{
&(d^{AIC} c^H_{~CF} f_{IDE} f_{HJ}^{~~~F} + d^{FIC} c_{HCJ} f_{IDE} 
f^{HA}_{~~~F} 
- d^{AIC} c_{HCJ} f_{IDF} f^{~HF}_E \cr &- d^{AIC} c_{HCJ} f_{IFE} f^{~HF}_D )
X^{a}_A X^{Eb} X^D \cdot X^J - (A \leftrightarrow E)~.
}}
Applying the Jacobi identity \dmnid{a}\ and the identity \dmnid{e}\ we
can reduce \secter\ to
\eqn\seclef{
(d^{AIC} c^H_{~CF} f_{IDE} f_{HJ}^{~~~F} - d^{FIA} c^H_{~CJ} f_{EDF}
f_{HI}^{~~~C} ) X^{a}_A X^{Eb} X^D \cdot X^J - (A \leftrightarrow E)~.
}
Using the symmetries and \dmnid{b}\ we find these terms cancel. This
proves no extra ordering terms arise and the classical computation of
\ezawa, goes through, with the understanding that $P$'s are ordered to 
the left. They find these terms in the commutator combine to give (see 
eqn. (88) \ezawa)
\eqn\eqeig{
\eqalign{
&-\frac{1}{2}d^{AIC}f_{I}^{\;DE}c_{HC}^{\quad F}f^{HBG}
\vec{P}_{B}\cdot \vec{X}_{G} (X^{a}_{A}X^{b}_{E}-X^{b}_{A}X^{a}_{E})
\vec{X}_{D}\cdot\vec{X}_{F}
\cr
& \quad -f^{DAE}f^{FBG}\vec{P}_{B}\cdot \vec{X}_{G}
(X^{a}_{A}X^{b}_{E}-X^{b}_{A}X^{a}_{E})
\vec{X}_{D}\cdot\vec{X}_{F},}
}
which are proportional to the purely bosonic part of the constraints.
The next step is to check that no extra terms are generated when this 
expression is symmetrically ordered. The two terms in \eqeig\ are
proportional to the second and third terms in \eigtwo\ and we have
already shown the ordering terms in these expressions vanish.
This completes the computation of the ordering terms that arise from
purely bosonic terms.

It remains to consider the fermionic terms.
The possible ordering terms that may be generated are of the form $X
\theta^2$ which arise from reordering $X^2 P \theta^2$ and $X
\theta^4$ terms.
Let us first consider the terms with four fermion operators. The
Clifford algebra identity \gammidd\ is required to simplify the second 
and third terms in eqn. (78) of \ezawa,
to express $[M^{-a}, M^{-b}]$ in terms of the constraints.
 We will
first argue that no ordering terms appear in the application of these
identities. Notice that no term of order $X$ can be
generated. The only possible terms would be of order $X \theta^2$.
To
see that these vanish notice that the relevant terms in the unordered
expansion of $[M^{-a}, M^{-b}]$ appear schematically as
\eqn\ftheta{
\theta_B \Gamma_1 \theta_C \theta_D \Gamma_2 \theta_E +
\theta_D \Gamma_2 \theta_E \theta_B \Gamma_1 \theta_C~,
}
where we have dropped the common prefactor, and $\Gamma_1$, $\Gamma_2$
are some products of the gamma matrices. This may be rearranged to
\eqn\fftheta{
\theta_B \Gamma_1 \theta_C \theta_D \Gamma_2 \theta_E +
\theta_E \Gamma_2 \theta_D \theta_C \Gamma_1 \theta_B~,
}
without generating any ordering terms. The ordering terms proportional
to $X \theta^2$ will then be proportional to the expression \fftheta\
with $\theta_C$ and $\theta_D$ contracted
\eqn\fthen{
\theta_B \Gamma_1 \Gamma_2 \theta_E + \theta_E \Gamma_2 \Gamma_1
\theta_B~.
}
Then
$\theta_E$ and $\theta_B$ can be reordered in the second term to
cancel the contribution from the first term. Thus we see no ordering
terms are generated in the application of the spinor identities to
the four fermion terms.

Now consider the terms that can appear in $A_1+A_2+A_3$. 
It suffices to check whether eqn. (86)
of \ezawa\ can pick up ordering terms. This equation is:
\eqn\eigsix{
\eqalign{
C^{(1)}&=\frac{i}{2}\{X^{a}_{A}(\theta^{A}\gamma^{b}\theta_{D})
{}-X^{b}_{A}(\theta^{A}\gamma^{a}\theta_{D})\}\varphi^{D}
\cr
& -\frac{i}{4}\{X^{a}_{A}(\theta_{B}\gamma^{b}\theta_{E})
{}-X^{b}_{A}(\theta_{B}\gamma^{a}\theta_{E})\}
d^{ABC} c_{DC}^{\quad E}\varphi^{D}
\cr
&+\frac{i}{2}(\theta_{D}\gamma^{abd}\theta^{D})X_{dE}\varphi^{E}
{}-\frac{i}{4}(\theta_{D}\gamma^{abd}\theta_{E})X_{dC}d_{A}^{\;DE}
c_{B}^{\;AC}\varphi^{B}, \cr}
}
which is to be understood as an unordered expression.
The ordering terms will be proportional to terms from
the contraction of $P$ with
the $X$'s and from contractions of pairs of $\theta$'s.
The terms from the first line of \eigsix\ cancel straightforwardly.
Using \dmnid{e}\ one likewise finds cancellation of terms on the
second line. For the first term on the third line of \eigsix\ the
contractions vanish trivially. For the second term on the third line
we use \ezaid{a}\ and \dmnid{b}\ to show cancellation of the terms.
This completes the proof that no ordering terms appear in
$A_1+A_2+A_3$.

Finally we must consider $A_4$. Let us just consider the second term
in \orterms\ 
and factor out $\theta_0 \gamma^b$.
The only terms in this expression that
can give us trouble are the $X^2 P \theta $ terms and the $X \theta^3$
terms which could give ordering terms of the form $X \theta$.
The computation reduces to showing the ordering terms between equations
(93) and (94) of \ezawa\ cancel. Eqn. (94) gives the $X \theta^3$
terms
\eqn\ninfou{
\eqalign{
(D_{4})_{\alpha} &=  \frac{i}{2}(\gamma^{a}\theta_{A})_{\alpha}
              (\theta_{B}\gamma\cdot
                  X_{C}\theta_{D}){d^{AB}}_{E}f^{CDE}\cr
        & +\frac{i}{4}(\gamma^{ad}\theta_{B})_{\alpha}X_{dA}
              (\theta_{C}\theta_{D})c^{ECD}{f^{AB}}_{E}\cr
        & +\frac{i}{4}\left[-(\gamma_{d}\theta_{A})_{\alpha}
                         (\theta_{B}\gamma^{ade}\theta_{C})
                       +(\gamma^{ed}\theta_{A})_{\alpha}
                         (\theta_{B}{\gamma^{a}}_{d}\theta_{C})\right]
                   X_{eD}{d^{BC}}_{E}f^{DAE} \cr
        & +\frac{i}{2}(\gamma_{d}\theta_{A})_{\alpha}
               (\theta_{B}\gamma^{d}\theta_{C})X^{a}_{D}
                 {d^{BD}}_{E}f^{ACE}~.\cr}
}
To simplify this equation, the Clifford algebra identity \gammidd\
must be applied to the third term in \ninfou.  To use this we must
show the ordering terms in this third term vanish. This follows
straightforwardly using the fact the $d_{ABC}$ is symmetric. To
further simplify \ninfou\ as in \ezawa\ we
must apply some group theory identities
to the remaining terms. Using the fact that $\Tr \gamma^a =0$ and
$c^A_{AB}=0$, it can be shown that no ordering terms are generated in
this process. We then can reduce $D_4$ to
\eqn\nineig{
\eqalign{
(D_{4})_{\alpha} &=
  {}-\frac{i}{4}(\gamma^{ad}\theta_{C})_{\alpha}X_{dD}
    (\theta_{B}\theta_{A})
    {c_{E}}^{CD}f^{EBA}
     \cr
    &-\frac{i}{3}\left[(\theta_{B}\theta_{A})\theta_{\alpha C}
            +(\theta_{C}\theta_{A})\theta_{\alpha B} \right]
        X^{a}_{D}{d^{BD}}_{E}f^{CAE} \cr
    & -\frac{i}{2}\left[(\gamma_{d}\theta_{[A})^{\alpha}
            (\theta_{B}\gamma^{d}\theta_{C]})\right]
            X^{a}_{D}{d^{BD}}_{E}f^{CAE}~.\cr}
}
The spinor identity \ffermid\ must be applied to the third
term in \nineig. Using the symmetries no ordering terms are generated
in this process. Combining the resulting expression with the $X^2 P
\theta$ terms we then obtain
\eqn\onehun{
[\tilde{Q}^{+}_{\alpha},\tilde{M}^{-a}]
=i\theta_{\alpha C}X^{a}_{D}{d^{CD}}_{E}\varphi^{E}
  +\frac{i}{2}(\gamma^{ad}\theta_{C})_{\alpha}X_{dD}
    {c_{E}}^{CD}\varphi^{E}~.
}
Finally we must check no ordering terms appear when this expression is 
(a)symmetrically ordered. The
terms generated by the first term vanish using the fact that $d_{ABC}$
is symmetric. The terms generated by the second term cancel when we
apply the identities \idstuff\ and \dmnid{b}.
This completes the proof that quantum ordering terms do not appear at
this order in $[M^{-a}, M^{-b}]$.

\subsec{ $[M^{-a}, H]$ }

The two possible nontrivial ordering contributions to this commutator
come from $[\tilde M^{-a}, \CM^2]$ and $[ \tilde Q^+, \CM^2]$.
Let us ignore the fermionic contributions for the moment and consider 
$[\tilde M^{-a}, \CM^2]$.
It is straightforward to see the only nontrivial
can appear
at order
$X^3$.
There are three ordering contributions:
\eqn\mahone{
\eqalign{
&[ \half d^{ABC} X_A^a P_B \cdot P_C, \half f_{DEF} f^D_{~GH}
X^E \cdot X^G X^F \cdot X^H ] \to \cr & \qquad \qquad
2 d^{ABC} f_B^{~DE} f_{CDG} (D-1)
X_A^a X_E \cdot X^G ~,\cr
&[ {1\over 4} d^{ABC} f_B^{~DE} f_C^{~FG} X_A^a X_D \cdot X_F X_E
\cdot X_G, P_H^2] \to  \cr & \qquad \qquad
-2 d^{ABC} f_B^{~DE} f_{CDG} (D-1) X_A^a X_E
\cdot X^G ~,\cr
&[ \half c^{ABC} P_A^a P_B \cdot X_C ,  \half f_{DEF} f^D_{~GH}
X^E \cdot X^G X^F \cdot X^H ] \to \cr & \qquad 2 ( c^{ABC} f_{DAF} f^D_{~BH} +
c^{AB}_{\quad F} f_{DAB} f^D_{~CH} + c^{AB}_{\quad H}
 f_{DAF} f^D_{~CB} ) X_C^a X^F \cdot
X^H ~.\cr}
}

The first two lines cancel against each other. Using the identities
$c^{ABF} f_{DAB} \sim \eta_{DF}$ and \dmnid{c},
one finds the third line vanishes. 
This completes the proof that the
ordering terms of bosonic origin vanish.

Now let us consider terms generated by the fermionic terms. Terms
of order $\theta^2$ cancel by symmetry considerations. $c$-number
terms cancel by $SO(9)$ invariance. The only non-trivial terms are
again of order $X^3$ would arise from reordering $X^3 \theta^2$
terms. This corresponds to properly ordering the terms in eqn.
(3.35d) of \dewit. The possible non-zero ordering contributions
are proportional to 
\eqn\mahfer{ \eqalign{ d^{ABC} f_A^{~FG}
f_C^{~DE} X_{Fb} X_{Gc} X_{Dd} \eta_{BE} \Tr( \gamma^{abc}
\gamma^d) ~,\cr 2 d^{ABC} f_C^{~DE} f_E^{~FG} X_A^a X_D^b X_F^c
\eta^{bc} \eta_{BG} ~.\cr} } 
The first term vanishes since the
$\Tr(\gamma^{abc} \gamma^d)=0$. The second term vanishes by
symmetry. The massaging of the properly ordered terms in the
commutator $[M^{-a}, H]$ using the identities of Appendix A, then
proceeds precisely as in \dewit. The final step is to show the
unordered expression (3.39) of \dewit\ does not give rise to ordering
terms. This is straightforward to verify, using the Jacobi identity
\dmnid{a}.

To complete the calculation we consider the commutator $[ \tilde Q^+,
\CM^2]$. The only problematic terms will come from ordering terms of
order $\theta^3$, which take the form
\eqn\tqter{
f_{ABC} \gamma_a \theta^A ( \theta^B \gamma^a \theta^C)~.
}
It is clear the ordering terms which arise by contractions of pairs of 
$\theta$'s will vanish by symmetry.
This completes the proof that the
ordering terms in $[M^{-a}, H]$ vanish at this order, 
as required by Lorentz invariance.

\newsec{Relation to Matrix Theory}

The key difference between the model of dWHN \dhn\ and Matrix theory \bfss\ 
is the treatment of
the longitudinal momentum. In dWHN this is identified with the
additional zero mode $P_0^+$ which satisfies a nontrivial commutation
relation with $q^-$. This mode is crucial for constructing the boost
generators of the previous sections. In Matrix theory, the size of the
matrices $N$ is identified with the longitudinal momentum 
$P^+ = N/R$ \foot{Here we have in mind the interpretation of
Matrix theory as the description of M-theory on a compact light-like
circle of radius $R$ \refs{\susskind,\seiberg}.} and the
challenge is to construct the conjugate to this operator which would
play the role of $q^-$ and allow us to construct boost
generators. 

For finite values of $N$ it seems difficult to construct such an
operator, as it maps between Matrix models with different values
of $N$. However progress can be made in the large $N$ limit. At
large $N$ we can consider a generalization of the ideas of Brezin
and Zinn-Justin \brezin, and construct a renormalization group equation
for the generating function of correlation functions $\CF$. (See
\sakai\ for further work on this approach and more recent references.) This
approach has been considered in the context of Matrix theory in
curved space by Douglas \douglas. The basic idea is to consider the
renormalization group flow when one row and one column of the
matrices is integrated out. In general, this should be equivalent
to a deformation of the theory with $N$ fixed. This may be
expressed as a renormalization group equation which takes the form
\eqn\renorm{ ( N {\partial \over \partial N} - \beta_i(\lambda_j) {\partial
\over \partial \lambda_i} +\gamma(\lambda_i)) \CF = r( \lambda_i) ~,
} 
where the
$\lambda_i$ are the coupling constants of
operators added to the action. 
For the case at hand it is difficult to carry out this integration
exactly, but the hope is that a systematic saddle point approximation
can be developed.

However,
if Matrix theory recovers eleven-dimensional Lorentz invariance in the
large $N$ limit (with $N/R$ fixed), 
this rescaling of $N$ corresponds to a longitudinal
boost. For Lorentz invariant physical quantities like S-matrix elements, the
renormalization group equation should simplify to
\eqn\simren{
 N {\partial \over \partial N} \CF =0~,
}
where it is assumed the parameters labeling the longitudinal momenta
are written as $P^+_i = \alpha_i N/R$, with $\alpha_i$ held fixed in
the derivative. 
For more general physical quantities $\CF$, the fact that 
$N$ should appear only in
the combination $N/R$ implies the renormalization group equation
takes the form
\eqn\simssren{
(N {\partial \over \partial N} + R {\partial \over \partial R})
\CF = 0 ~.
}
This equation can then be thought of as 
a definition for the conjugate of $N$, in terms of a derivative with
respect to $R$.

%Let us see how we can use this idea to relate the dWHN and BFSS models.
%Notice that by rescaling the time coordinate $t$,
%the dWHN Hamiltonian \sphham\ can be written in a form independent of
%$P_0^+$, and we obtain the Matrix theory Hamiltonian
%\eqn\hammat{
%H  =  \frac{\vec{P}_0^2+\CM^2}{2}
%}
%The eigenstates
%of this Hamiltonian will contain no explicit dependence on $R$, the
%radius of the compactified light-like direction. Formally we can then
%covariantize the states of Matrix theory by multiplying them by the
%$R$ dependent phase
%\eqn\rphase{
%| \psi \rangle \to e^{i \frac{N}{R} x_0^-} | \psi \rangle
%}

The results for the Lorentz algebra of the previous section may 
then be carried over to Matrix 
theory by replacing $P_0^+$ with $N/R$ and 
$q^-$ with
\eqn\bfssqm{
q^- \to -i {R^2\over N } {\partial\over  \partial R}~.
}
Related ideas in the context of a membrane in four dimensions have
been discussed in \jevicki.
The real challenge remaining is to verify \simssren\ directly by performing an
integration over a row and a column in the large $N$ limit.

The definition for $X^-_A$ \xiym\ gives us the
non-abelian version of the longitudinal coordinate as 
an operator written in terms of the
transverse degrees of freedom and $R$. Assuming the above proposal is
correct, it now
becomes possible to construct states in Matrix theory localized in the
longitudinal direction.

\bigskip
{\bf Acknowledgments}

I wish to thank M. Douglas, A. Jevicki and W. Taylor for helpful discussions.
I thank the IAS and ITP, Santa Barbara for hospitality during
the course of this research.
This research is supported in part by DOE grant DE-FE0291ER40688-Task A.     
 
\vfil
\eject
%and
%are proportional to
%\eqn\mhterms{
%\eqalign{
%d_{ABC} f^B_{DE} f^{DC}_G X_A^a X_E \cdot X_G \cr
%d_{ABC} f^{BA}_E f^C_{FG} X_F^a X_E \cdot X_G \cr
%d_{ABC} f^{B}_{DE} f^{CD}_G X_A^a X_E \cdot X_G \cr
%c_{ABC} f^B_{DE} f^{DC}_G X_A^a X_E \cdot X_G \cr}
%}

%The second line vanishes by symmetry.
%Instead, look at $X^3$ term in $[M^{-a}, H]$. This does not vanish by
%any obvious
%symmetry considerations.

\appendix{A}{Area preserving diffeomorphism identities}

The tensors $f_{ABC}$, $d_{ABC}$ and $c_{ABC}$ satisfy a number of
identities in the large $N$ limit, up to $1/N^2$ corrections.
These are derived using the completeness relations \sphcom, symmetry,
and integration by parts \dewit\
\eqna\dmnid
$$
\eqalignno{
{f_{[AB}}^{E}f_{C]DE} &= 0 ,& \dmnid a\cr
c_{ABC}+c_{ACB}
&= 2\int d^{2}\sigma \sqrt{w(\sigma)}\frac{1}{\omega}_{A}
    \Delta Y_{A}(\sigma)Y_{B}(\sigma)Y_{C}(\sigma)
=-2d_{ABC},& \dmnid b\cr
{c_{DE}}^{[A}f^{BC]E}
&=2\int d^{2}\sigma \sqrt{w(\sigma)}
  \frac{w^{rs}(\sigma)}{\omega_{D}}\partial_{r}Y_{D}(\sigma)
  \partial_{s}Y^{[A}(\sigma)
  \{ Y^{C}(\sigma),Y^{B](\sigma)}\}
=0 ,& \dmnid c\cr
d_{ABC}{f^{A}}_{[DE}{f^{B}}_{F]G}
&=\int d^{2}\sigma \sqrt{w(\sigma)}
  Y_{C}(\sigma)\{Y_{[D}(\sigma),Y_{E}(\sigma)\}
  \{ Y_{F]}(\sigma),Y_{G}(\sigma)\}
=0,& \dmnid d\cr
{f_{A(B}}^{E}d_{CD)E}
&=\int d^{2}\sigma \sqrt{w(\sigma)}
  \frac{1}{3}
  \{Y_{A}(\sigma)\,,\,Y_{B}(\sigma)Y_{C}(\sigma)Y_{D}(\sigma)\}
= 0 ,& \dmnid e\cr
d_{EA[B}{d_{C]D}}^{E}
&=\int d^{2}\sigma \sqrt{w(\sigma)}
  Y_{A}Y_{[B}Y_{C]}Y_{D}
  {}-\displaystyle\int d^{2}\sigma \sqrt{w(\sigma)}
    Y_{A}Y_{[B}
  \displaystyle\int d^{2}\rho \sqrt{w(\rho)}
    Y_{C]}Y_{D}\cr
&= -\eta_{A[B}\eta_{C]D}.& \dmnid f\cr}
$$

For the special case of spherical topology, \dewit\ found
\eqn\idstuff{
{f_{AB}}^{E}c_{ECD}
  = c_{EAB}{f^{E}}_{CD}-2{f_{BD}}^{E}d_{ACE}~.
}

Ezawa et al. \ezawa\ extended these identities to include
\eqna\ezaid
$$
\eqalignno{
-c^{ABC}{d_{C}}^{EF}+2c^{AC(E}{d_{C}}^{F)B}
   &= 4\eta^{A(E}\eta^{F)B}, & \ezaid a\cr
\frac{1}{4}({c_{C}}^{FE}c^{[AB]C}+{c^{[A|F|}}_{C}c^{B]CE})
      &= -\frac{1}{2}\left(\frac{1}{\omega_{A}}-\frac{1}{\omega_{B}}
                    \right)
         \frac{1}{\omega_{C}}{f_{C}}^{EF}f^{CAB} \cr & \qquad
       +\frac{1}{2\omega_{A}\omega_{B}}{f_{C}}^{AB}f^{CEF}, &\ezaid b
     \cr
{d_{E}}^{AC}(c^{DEB}-2d^{DEB})-2c^{D(C|E|}{d_{E}}^{A)B}
    &=4\eta^{AC}\eta^{BD}, &\ezaid c\cr
{d^{CG}}_{H}{d^{DH}}_{I}f^{I[EF}{f^{B]A}}_{G}
   &=-f^{D[EF}f^{B]AC}. &\ezaid d\cr}
$$

\appendix{B}{$SO(9)$ Clifford Algebra Identities}

We review here some definitions and identities that are used
above. The gamma matrices ${\gamma^{a}}_{\alpha\beta}$
($a=1,\ldots ,9$ ; $\alpha , \beta =1,\ldots ,16$) are taken to be
real and symmetric matrices. Using these gamma matrices we can
construct an orthogonal complete basis for $16\times 16$ real
matrices \eqn\gambas{ \left\{ I_{\alpha\beta}\ ,
       \ {\gamma^{a}}_{\alpha\beta}\ ,
       \ {\gamma^{ab}}_{\alpha\beta}\ ,
       \ {\gamma^{abc}}_{\alpha\beta}\ ,
       \ {\gamma^{abcd}}_{\alpha\beta}\ \right\},
}
where we have defined
\eqn\stiff{
\gamma^{a_{1}\cdots a_{k}}
   =\gamma^{[a_{1}}\gamma^{a_{2}}\cdots \gamma^{a_{k}]}.
}
$I$, $\gamma^{a}$ and $\gamma^{abcd}$ are
symmetric, and $\gamma^{ab}$ and $\gamma^{abc}$ are antisymmetric
with respect to the spinorial indices.

Some useful identities are:
\eqn\gammid{
\gamma^{a}\gamma^{b_{1}\cdots b_{k}}
  =\gamma^{ab_{1}\cdots b_{k}}
     +\sum_{l=1}^{k}(-)^{l-1}\delta^{ab_{l}}
             \gamma^{b_{1}\cdots\check{b}_{l}\cdots b_{k}}\ \ ,
}
\eqn\gammidd{
\eqalign{
(\gamma^{b})_{\alpha\beta}(\gamma_{ab})_{\gamma\delta}
   +(\gamma^{b})_{\gamma\delta}(\gamma_{ab})_{\alpha\beta}
+(\gamma^{b})_{\alpha\delta}(\gamma_{ab})_{\gamma\beta}
+(\gamma^{b})_{\gamma\beta}(\gamma_{ab})_{\alpha\delta} \cr
-2I_{\delta\beta}(\gamma_{a})_{\gamma\alpha}
+2I_{\alpha\gamma}(\gamma_{a})_{\beta\delta}=0 ~.}
}

By multiplying \gammidd\ by $(\gamma^{a})_{\delta\epsilon}
  (\theta_{[A}^{\beta}\theta_{B}^{\gamma}\theta_{C]}^{\epsilon})$ we
obtain
\eqn\ffermid{
(\gamma_{d}\theta_{[A})^{\alpha}(\theta_{B}\gamma^{d}\theta_{C]})
  =\theta^{\alpha}_{[A}(\theta_{B}\theta_{C]})\ \ .
}

\listrefs
\end